\documentclass{pj}
\usepackage{graphicx}

\usepackage[]{cite}

\usepackage{tikz}

\usepackage[fleqn]{amsmath}
\usepackage[]{amssymb}
\usepackage[]{mathrsfs}
\usepackage[]{eucal}
\usepackage[]{tensor}
\usepackage[]{amsthm}
\usepackage[]{bm}

\newcommand{\defeq}{\mathrel{\mathop:}=}

\newcommand{\dd}{\partial}
\newcommand{\nab}[1]{\nabla_{\! #1}}
\DeclareBoldMathCommand{\bnab}{\nabla}

\newcommand{\vol}{\df V}

\newcommand{\norm}[1]{\left\lVert #1 \right\rVert}

\newcommand{\df}{\mathrm{d}}

\newcommand{\qqd}{\ , \quad}
\newcommand{\bc}{\begin{center}}
\newcommand{\ec}{\end{center}}
\newcommand{\be}{\begin{equation}}
\newcommand{\ee}{\end{equation}}

\def\bal#1\eal{\begin{align}#1\end{align}}

\newcommand{\mx}[1]{\bm{\mathsf{#1}}}
\newcommand{\trans}{^{\mathsf{T}}}

\newcommand{\vct}[1]{\mathbf{#1}}

\newcommand{\0}{\vct{0}}

\usepackage{dsfont}
\newcommand{\rr}{\mathds{R}}
\newcommand{\cl}[1]{\overline{#1}}

\usepackage[unicode]{hyperref}
\usepackage{xcolor}
\definecolor{pastblue}{HTML}{336699}
\hypersetup{colorlinks,linkcolor={pastblue},citecolor={pastblue},urlcolor={pastblue}}

\theoremstyle{plain} \newtheorem{tm}{Theorem}[section]
\theoremstyle{plain} \newtheorem{lm}[tm]{Lemma}
\theoremstyle{definition} \newtheorem{defn}[tm]{Definition}
\newcommand{\btm}{\begin{tm}}
\newcommand{\etm}{\end{tm}}
\newcommand{\blm}{\begin{lm}}
\newcommand{\elm}{\end{lm}}
\newcommand{\bdefn}{\begin{defn}}
\newcommand{\edefn}{\end{defn}}

\begin{document}

\setcounter{page}{1}
\pjheader{Vol.\ x, y--z, 2021}

\title[Capacitance matrix revisited]
{Capacitance matrix revisited}
 \footnote{\it Received date}  \footnote{\hskip-0.12in*\, Corresponding
author:~Ivica~Smoli\'c (ismolic@phy.hr).}
\footnote{\hskip-0.12in\textsuperscript{1} Department of Physics, Faculty of Science, University of Zagreb, Bijeni\v cka cesta 32, 10000 Zagreb, Croatia.}

\author{Ivica~Smoli\'c\textsuperscript{*, 1} and Bruno~Klajn\textsuperscript{1}}

\runningauthor{Smoli\'c and Klajn}

\tocauthor{Ivica~Smoli\'c and Bruno~Klajn}

\begin{abstract}
The capacitance matrix relates potentials and charges on a system of conductors. We review and rigorously generalize its properties, block-diagonal structure and inequalities, deduced from the geometry of system of conductors and analytic properties of the permittivity tensor. Furthermore, we discuss alternative choices of regularization of the capacitance matrix, which allow us to find the charge exchanged between the conductors having been brought to an equal potential. Finally, we discuss the tacit approximations used in standard treatments of the electric circuits, demonstrating how the formulae for the capacitance of capacitors connected in parallel and series may be recovered from the capacitance matrix.
\end{abstract}

%\mytableofcontents
%\tableofcontents

\setlength {\abovedisplayskip} {6pt plus 3.0pt minus 4.0pt}
\setlength {\belowdisplayskip} {6pt plus 3.0pt minus 4.0pt}

\

%%%%%%%%%%%%%%%%%%%%%%%%%%%%%%
%%%%%%%%%%%%%%%%%%%%%%%%%%%%%%
\section{Introduction} %%%%%%%
%%%%%%%%%%%%%%%%%%%%%%%%%%%%%%
%%%%%%%%%%%%%%%%%%%%%%%%%%%%%%

One of the fundamental problems of electrostatics considers the system of charged ideal conductors. In a basic setting one either assigns the potentials on each of the conductors and asks for a total charge on each of them or, vice versa, asks for potentials if the charges are known. Since Maxwell's equations are linear, this relation turns out also to be  linear, and given by corresponding matrices: The \emph{capacitance matrix} produces charges from potentials, while the \emph{potential matrix} solves the opposite problem. Hybrid problems in which potentials are fixed on some conductors, while total charges are prescribed on others, may be treated with the same set of tools.

\smallskip

The origin of the capacitance matrix goes back at least to Maxwell's \emph{Treatise} \cite{Maxwell}, where it was introduced and some of its basic properties derived\footnote{Maxwell refers to elements of the capacitance matrix as ``coefficients of induction'', while the elements of the potential matrix are referred to as the ``potential coefficients''.}. Several subsequent textbooks on classical electrodynamics, such as Smythe \cite{Smythe}, Landau and Lifshitz \cite{LLED}, and Jackson \cite{Jack}, contain only brief discussions about the capacitance matrix, while similar (or even briefer) analysis may be found in numerous more recent references \cite{CPK,Schwartz,Wangsness,NB,Ohanian,Greiner,PP,Muller,Vanderlinde,Zangwill,Garg,Toptygin,Durand}. A slightly more detailed treatment of the problem is presented in \cite{SDeRMTN} (section 24.6), which makes use of the Green's function and generalizes basic results in presence of isotropic inhomogeneous dielectric. A nice alternative discussion about the properties of the capacitance matrix can also be found in \cite{HD08}, which was later formalized in \cite{HD11}.

\smallskip

Still, the situation in the literature can hardly be described as satisfactory, both from formal and practical points of view. Published proofs rely on many tacit assumptions and suffer from various technical gaps. This is not just a mere nitpicking, as one of the elementary problems, determination of the charge exchanged between the conductors having been brought to equal potential, relies on some subtle mathematical details. Furthermore, one could argue that the actual use of the capacitance matrix is somewhat obscured in textbooks and its conspicuous absence in treatments of the electric circuits with capacitors calls for an elaboration. In order to remedy these deficiencies, we shall thoroughly review the properties of the capacitance matrix, with particular focus on the delicate issue of its regularization, and discuss its use in several concrete problems. 

\smallskip

The paper is organized as follows. In section 2 we give a precise definition of the problem, including geometric conditions on the domain and algebraic and analytic conditions on the permittivity tensor. In sections 3 and 4 we introduce the capacitance matrix (using the auxiliary Dirichlet problem) with some brief remarks on problems with unbounded conductors. Section 5 is devoted to the detailed study of the properties of the capacitance matrix, in which we extend some previous results from the literature. In section 6 we discuss several procedures for the regularization of the capacitance matrix and in section 7 we give an example how a regularized capacitance matrix may be utilized in a concrete physical problem. In section 8 we discuss various approximations which are tacitly used in the analysis of the electric circuits and explain how one can recover formulae for the total capacitance of the capacitors connected in parallel and series from the elements of the capacitance matrix. Finally, in section 9 we give concluding remarks, while in appendix A we give a brief overview of the existence and uniqueness results for the Dirichlet problem and a proof of one important technical theorem. 

\medskip

\emph{Technical remarks}. For any set $A$ we denote its interior by $A^\circ$, boundary by $\dd A$, closure by $\cl{A}$, and if $B$ is any other set, their difference by $A - B$. An open ball centred at $x \in \rr^3$, with radius $r > 0$, is denoted by $B(x,r)$. The exterior cone condition is defined in the Appendix A. Matrices are denoted with bold symbols, e.g.~$\mx{A}$.

%%%%%%%%%%%%%%%%%%%%%%%%%%%%%%%%%%%%%%
%%%%%%%%%%%%%%%%%%%%%%%%%%%%%%%%%%%%%%
\section{System of conductors} %%%%%%%
%%%%%%%%%%%%%%%%%%%%%%%%%%%%%%%%%%%%%%
%%%%%%%%%%%%%%%%%%%%%%%%%%%%%%%%%%%%%%

The question that introduces the capacitance matrix starts from potentials and charges on (topologically) connected components of the system of ideal conductors. Nevertheless, we find that for many technical reasons it is much better to focus the beginning of the discussion on the complement of conductors and components of its boundary. 

\smallskip

\begin{figure}[!ht]
\centering
\begin{tikzpicture}
\draw[thick,gray,fill=lightgray] (0,0) ellipse (2.5 and 1.75);
\draw[thick,gray,fill=white] (-0.5,0) circle (0.7);
\node at (1.2,0) {\small $K_1$};
\node at (-0.5,0) {\small $\Omega_1$};
\node at (-1.83,0.83) {\small $\mathcal{S}_1$};
\node at (-1.0,-0.7) {\small $\mathcal{S}_2$};
\node at (4.2,0) {\small $\Omega_{\mathrm{e}}$};
\draw[thick,gray,fill=lightgray] (6,0) [out=90,in=180] to (7,2) [out=0,in=120] to (9,0.5) [out=60,in=180] to (10,1) [out=0,in=90] to (11,0) [out=-90,in=0] to (8.5,-2.5) [out=180,in=-90] to (6,0);
\draw[thick,gray,fill=white] (7.2,0.75) circle (0.7);
\draw[thick,gray,fill=white] (8.5,-1.2) ellipse (1.5 and 0.8);
\draw[thick,gray,fill=lightgray] (8.2,-1.2) circle (0.5);
\node at (8.2,-1.25) {\small $K_3$};
\node at (7.2,0.75) {\small $\Omega_2$};
\node at (9.3,-1.2) {\small $\Omega_3$};
\node at (10,0) {\small $K_2$};
\node at (6.25,-1.5) {\small $\mathcal{S}_3$};
\node at (6.7,0.05) {\small $\mathcal{S}_4$};
\node at (8.5,-0.2) {\small $\mathcal{S}_5$};
\node at (7.5,-1.25) {\small $\mathcal{S}_6$};
\end{tikzpicture}
\caption{Schematic representation (cross-section) a system of conductors with $(M,N,P) = (4,6,3)$. Note that $\dd K_1 = \mathcal{S}_1 \cup \mathcal{S}_2$, $\dd K_2 = \mathcal{S}_3 \cup \mathcal{S}_4 \cup \mathcal{S}_5$ and $\dd K_3 = \mathcal{S}_6$; also $\dd\Omega_{\mathrm{e}} = \mathcal{S}_1 \cup \mathcal{S}_3$, $\dd\Omega_1 = \mathcal{S}_2$, $\dd\Omega_2 = \mathcal{S}_4$ and $\dd\Omega_3 = \mathcal{S}_5 \cup \mathcal{S}_6$. The Dirichlet boundary conditions for $\mathcal{S}_1$ and $\mathcal{S}_2$ have to be the same and likewise for $\mathcal{S}_3$, $\mathcal{S}_4$ and $\mathcal{S}_5$.}
\label{fig:sc}
\end{figure}
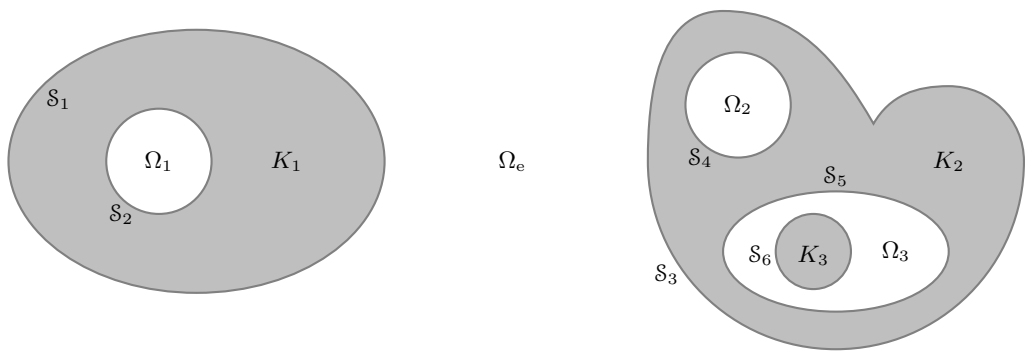

We assume that the system of conductors $K \subseteq \rr^3$ is a closed set, consisting of a finite number of connected components $\{K_1,\dots,K_P\}$, each of which physically represents an individual conductor. As any connected conducting object is held at a constant potential, the number of independent potentials in the problem is $P$. Furthermore, since physically meaningful observables are potential \emph{differences}, the effective reduced space of potentials is $\rr^{P-1}$. The domain on which we analyze the problem is the open set $\Omega \defeq \rr^3 - K$, which consists of a finite number of connected components, $\{\Omega_1,\dots,\Omega_M\}$, representing cavities and/or space between the conductors. In particular, if there is an unbounded connected component of $\Omega$, we denote it additionally by $\Omega_{\mathrm{e}}$ (here ``e'' stands for the ``external''), with $\mathrm{e} \in \{1,\dots,M\}$. The boundary of the domain, $\dd\Omega$, consists of a finite number of connected components, $\{\mathcal{S}_1,\dots,\mathcal{S}_N\}$, which are simultaneously physical boundaries of conductors. On each of them a constant Dirichlet condition must be imposed (as in general we have $N \ge P$, these boundary conditions will not be independent). An example of a system of conductors is illustrated in Figure \ref{fig:sc}.

\smallskip

This setting is still too general and one must narrow down the specifications on the geometry of the problem in order to make some progress.

\smallskip

\bdefn
We say that a nonempty open set $\Omega \subseteq \rr^3$ is \emph{basic} if it satisfies the following conditions:
\begin{itemize}
\item[(1)] both $\Omega$ and its complement $\rr^3 - \Omega$ have finite number of connected components;

\item[(2)] $\Omega$ is either bounded or it has one unbounded connected component, $\Omega_{\mathrm{e}}$, such that its complement $\rr^3 - \Omega_{\mathrm{e}}$ is a compact set; 

\item[(3)] all connected components of the complement $\rr^3 - \Omega$ have nonempty interior;

\item[(4)] closures of connected components of $\Omega$ are pairwise disjoint and connected components of the complement $\rr^3 - \Omega$ are pairwise disjoint;

\item[(5)] $\Omega$ has a compact, orientable, piecewise smooth boundary $\dd\Omega$, and the boundary of each connected component of $\Omega$ satisfies the exterior cone condition.
\end{itemize}
\edefn

\smallskip

These particular choices deserve a brief justification. Here we leave aside physically unrealistic cases with (1) infinite number of conductors or cavities, (2) cases in which both the conductor and its complement are unbounded, and (3) measure zero conductors, such as a system which contains a single conducting point as one of its connected components. Condition (3), however, excludes some simple textbook examples of conductors (such as a conducting 2-sphere) which, nevertheless, can in principle always be simply defined as the limit of the basic ones. Condition (4), as a mere technical simplification, excludes cases where either two parts of the domain $\Omega$ or two pieces of conductor $K$ ``touch'' (e.g.~two cavities in a conducting bulk, separated by a single conducting point), all of which are again just limits of basic cases. Finally, the assumption (5) picks out conductors with boundary which is not too ``irregular'' (say, a fractal) so that it is suitable for simple manipulations of integrals with Stokes' theorem \cite{LeeSmooth,Federer,Morgan} and simultaneously satisfies a sufficient condition that guarantees the existence of the solution of the associated Dirichlet problem (see discussion in Appendix A). In the rest of the paper we shall assume that $\Omega$ is a basic open set, as defined above.

\smallskip

%%% https://mathoverflow.net/questions/253488/stokes-theorem-for-manifolds-with-boundary-as-disjoint-union-of-submanifolds
%%% https://math.stackexchange.com/questions/469437/what-is-the-boundary-in-stokes-theorem

We note in passing that numbers $M$, $N$ and $P$ satisfy the constraint $M+P=N+1$. This relation is trivially satisfied in the absence of conductors, where $M=1$ and $P=N=0$, while addition of each new hypersurface $\mathcal{S}_i$, satisfying the conditions from the definition above, increases $N$ by $1$ and simultaneously increases $M+P$ by $1$. Namely, our system of conductors may be built inductively, starting from the empty space $\rr^3$, in which we place pieces of conductors and drill cavities in them. Each step involves the introduction of a novel compact hypersurface in an open ball, either $B \subseteq \Omega$ or $B \subseteq K$, which according to Jordan--Brouwer separation theorem (see \cite{GP}, section 2.5, and \cite{Lima88,McGrath16} for smooth version; \cite{PMK09} for polyhedral generalization), divides the ball into ``outside'' and ``inside''. If the introduction of this hypersurface results with a new conductor then $(\Delta M,\Delta P) = (0,1)$, whereas if the result is a new cavity in conductor, then $(\Delta M,\Delta P) = (1,0)$.

\medskip

\begin{figure}[!ht]
\centering
\begin{tikzpicture}
\draw[thick,gray,fill=lightgray,rounded corners=2pt] (-1.1,-1.1) rectangle (3.2,1.1);
\draw[thick,gray,fill=white] (-0.9,-0.9) rectangle (3.0,0.9);

\draw[thick,gray,fill=lightgray] (0,0) circle (0.5);
\draw[thick,thick,gray,fill=white] (0,0) circle (0.3);
\draw[thick,thick,gray,fill=lightgray] (0,0) circle (0.1);

\draw[thick,gray,fill=lightgray] (1.7,0) ellipse (0.8 and 0.6);
\draw[thick,gray,fill=white] (1.7,0) ellipse (0.6 and 0.5);
\draw[thick,gray,fill=lightgray] (1.5,0) circle (0.1);
\draw[thick,gray,fill=lightgray] (1.9,0) circle (0.1);

\draw[thick,gray,fill=lightgray] (4.7,0) ellipse (0.7 and 0.8);
\draw[thick,gray,fill=white] (4.7,0) ellipse (0.6 and 0.5);
\draw[thick,gray,fill=lightgray] (4.4,0) circle (0.1);
\draw[thick,gray,fill=lightgray] (4.9,0) circle (0.15);

\begin{scope}[shift={(8,-1.75)}] %%% conductor tree
\draw[black,fill=black] (1,3) circle (0.05);
\draw[black,fill=black] (0,2) circle (0.05);
\draw[black,fill=black] (2,2) circle (0.05);

\draw[black,fill=black] (-0.5,1) circle (0.05);
\draw[black,fill=black] (0.5,1) circle (0.05);
\draw[black,fill=black] (1.5,1) circle (0.05);
\draw[black,fill=black] (2.5,1) circle (0.05);

\draw[black,fill=black] (0.2,0) circle (0.05);
\draw[black,fill=black] (0.8,0) circle (0.05);
\draw[black,fill=black] (-0.5,0) circle (0.05);

\draw[thin] (1,3) -- (0,2);
\draw[thin] (1,3) -- (2,2);
\draw[thin] (0,2) -- (-0.5,1);
\draw[thin] (0,2) -- (0.5,1);
\draw[thin] (2,2) -- (1.5,1);
\draw[thin] (2,2) -- (2.5,1);
\draw[thin] (0.5,1) -- (0.2,0);
\draw[thin] (0.5,1) -- (0.8,0);
\draw[thin] (-0.5,1) -- (-0.5,0);
\end{scope}

\end{tikzpicture}
\caption{Left: Schematic representation (cross-section) a system of conductors with $(M,N,P) = (5,13,9)$. Right: The associated conductor tree.}
\label{fig:ct}
\end{figure}
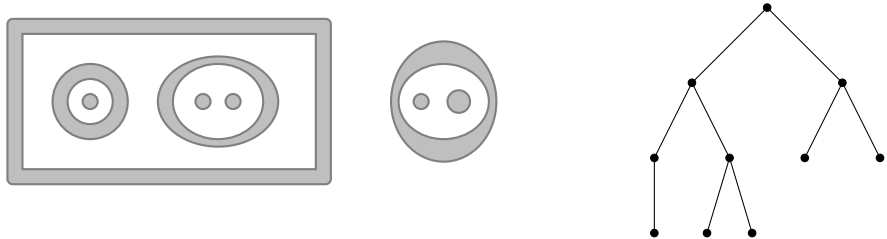

There is a convenient way to represent the topology of conductors involved in the problem using graph theory. Each conductor is represented by a vertex and a pair of vertices is connected by a line (edge) if they correspond to a pair of conductors sharing a boundary with a common connected component of $\Omega$ and if one of them bounds the other. In addition, if $K$ is bounded, one vertex is assigned to the ``point at infinity'' and connected to all the conductors sharing a boundary with $\Omega_\mathrm{e}$. The result of this procedure is in general a \emph{tree}, a connected undirected acyclic graph, which in this context might be called a \emph{conductor tree}. One can say that each edge represents a potential difference, so that number of edges is $P$. An example, with $(M,N,P) = (5,13,9)$, is sketched in Figure \ref{fig:ct}.

\smallskip

The domain $\Omega$ is, for generality, assumed to be filled with linear dielectric material with local response which, as a special case, may be vacuum. Linear dielectric is a medium in which the electric scalar potential $\Phi$ is a solution of the homogeneous partial differential equation (free charge is placed on the boundary $\dd\Omega$)
\be\label{eq:hardie}   %%% "harmonic dielectric"
\nab{a} (\hat{\epsilon}^{ab} \, \nab{b}\Phi) = 0 \ ,
\ee
where $\hat{\epsilon}_{ab}$ is the relative permittivity tensor (we use hatted epsilon in order to avoid an additional index ``r''). For example, in the vacuum case we have $\tensor{\hat{\epsilon}}{^a_b} = \tensor{\delta}{^a_b}$. The corresponding dimensionful permittivity tensor is $\epsilon_{ab} = \epsilon_0 \, \hat{\epsilon}_{ab}$. Again, the problem with completely arbitrary tensor $\hat{\epsilon}_{ab}$ would be too difficult to treat, so we make some additional assumptions. 

\smallskip

\bdefn
The relative permittivity tensor $\hat{\epsilon}_{ab}(x)$ is a $C^1$ tensor field such that
\begin{itemize}
\item[(a)] it is symmetric, $\hat{\epsilon}_{ab} = \hat{\epsilon}_{ba}$, and

\item[(b)] it is bounded in a sense that there exists a real constant $\kappa \ge 1$, such that the following inequalities
\be\label{eps}
v_c w^c \le \hat{\epsilon}_{ab}(x) v^a w^b \le \kappa v_c w^c
\ee
hold for all $x \in \Omega$ and all nonzero vectors $v^a$ and $w^a$ at $x$.
\end{itemize}
\edefn

\smallskip

\noindent
Equation (\ref{eq:hardie}), with the relative permittivity tensor satisfying the conditions above, belongs to a class of linear uniformly elliptic partial differential equations. Namely, condition (b) is equivalent to the assumption that $\lambda_i(x) \in [1,\kappa]$ for all eigenvalues $\lambda_i(x)$ of $\hat{\epsilon}_{ab}$ at each point of the domain $x \in \Omega$ (see e.g.~\cite{GT}, p.~1). Although the ellipticity itself could be assured by a weaker assumption, such as $0 < \hat{\epsilon}_{ab}(x) v^a w^b$, we need condition (b) for the existence results and some inequalities between the elements of the capacitance matrix. We shall not consider here cases of relative permittivity tensor with lower order differentiability, such as a piecewise smooth $\hat{\epsilon}_{ab}$ for a capacitor filled with layers of different dielectrics. Also, we stress that in some of the results discussed below we need much stronger smoothness assumption for sharper inequalities, namely that $\hat{\epsilon}_{ab}$ is real analytic (this will be clearly emphasized).

\smallskip

From a physical standpoint, the upper bound $\hat{\epsilon}_{ab}(x) v^a w^b \le \kappa v_c w^c$ follows from the assumption that dielectric response will not be unbounded on a domain. The two other conditions may be justified using thermodynamic arguments: in order to see (a) we express the components of the permittivity tensor as second derivatives of the corresponding Helmholtz free energy (see e.g.~page 54 in \cite{LLED} and section 33 in \cite{Kittel}), while the inequality $v_c w^c \le \hat{\epsilon}_{ab}(x) v^a w^b$ follows from demand that the total Helmholtz free energy is bounded from below (see e.g.~page 59 in \cite{LLED}).

%%%%%%%%%%%%%%%%%%%%%%%%%%%%%%%%%%%%%%%%%%%%%
%%%%%%%%%%%%%%%%%%%%%%%%%%%%%%%%%%%%%%%%%%%%%
\section{Auxiliary Dirichlet problem} %%%%%%%
%%%%%%%%%%%%%%%%%%%%%%%%%%%%%%%%%%%%%%%%%%%%%
%%%%%%%%%%%%%%%%%%%%%%%%%%%%%%%%%%%%%%%%%%%%%

A systematic approach to the properties of the capacitance matrix, makes use of the auxiliary elementary solutions: for each $i \in \{1,\dots,N\}$ we define $u_i$ as a solution of (\ref{eq:hardie}) with Dirichlet boundary condition $u_i|_{\mathcal{S}_j} = \delta_{ij}$ and asymptotic condition $\lim_{\norm{x}\to\infty} u_i(x) = 0$ on $\Omega_{\mathrm{e}}$ (if there is an exterior region of the domain $\Omega$). In other words, $u_i$ is a ``potential'' which forms when we fix the unit potential on the surface $\mathcal{S}_i$, and ground the rest of the surfaces. Note, however, that $u_i$ in general \emph{does not} represent physical potential, as some surfaces on a boundary of the same conductor here might have different values of potential ($0$ and $1$). For example, if we have a single conductor $K$ filling up the space between two concentric spheres, where the inner sphere is denoted by $\mathcal{S}_1$ and the outer by $\mathcal{S}_2$, then $u_1 = 1$ on $\mathcal{S}_1$ and $u_1 = 0$ on $\mathcal{S}_2$, representing a nonphysical situation in which the potential is not constant on an ideal connected conductor. Foundational questions about the existence and uniqueness of the auxiliary functions $u_i$ are discussed in detail in the Appendix A.

\smallskip

In a general Dirichlet problem we have a set of potential values $\{\varphi_1,\dots,\varphi_N\} \subseteq \rr$ given on boundary hypersurfaces $\{\mathcal{S}_1,\dots,\mathcal{S}_N\}$. In addition, if $\Omega$ is unbounded, we demand asymptotic condition that $\Phi(x) \to \varphi_\infty$ as $\norm{x} \to \infty$ for some $\varphi_\infty \in \rr$ (usual physical choice is $\varphi_\infty = 0$). If $\Omega$ is bounded we may again denote the value of the potential $\Phi$ on outermost $\mathcal{S}_i$ by $\varphi_\infty$. Due to linearity of the equation (\ref{eq:hardie}), the solution is simply given by superposition
\be\label{eq:Phisum}
\Phi(x) = \varphi_\infty + \sum_{i=1}^N (\varphi_i - \varphi_\infty) u_i(x) \ . 
\ee
A tacit assumption here is that the values of potentials $\varphi_i$ on each component of the boundary $\dd K_p$ of a connected conductor $K_p$ are equal, as the boundary of a connected conductor is an equipotential surface.

%%%%%%%%%%%%%%%%%%%%%%%%%%%%%%%%%%%%
%%%%%%%%%%%%%%%%%%%%%%%%%%%%%%%%%%%%
\section{Capacitance matrix} %%%%%%%
%%%%%%%%%%%%%%%%%%%%%%%%%%%%%%%%%%%%
%%%%%%%%%%%%%%%%%%%%%%%%%%%%%%%%%%%%

Let us denote by $n^a$ normal to $\dd\Omega$, pointing \emph{into} the $\Omega$. Gauss law implies that the total free charge on $i$-th surface is given by the integral
\be
Q_i = -\oint_{\mathcal{S}_i} n^a \, \tensor{\epsilon}{_a^b} (\nab{b} \Phi) \, \df a \ .
\ee
If for potential $\Phi$ we insert the sum (\ref{eq:Phisum}), this allows us to write
\be
Q_i = \sum_{j=1}^N C_{ij} (\varphi_j - \varphi_\infty)
\ee
where we have introduced the \emph{capacitance matrix} $\mx{C}$ with elements
\be\label{eq:basicC}
C_{ij} \defeq -\oint_{\mathcal{S}_i} n^a \, \tensor{\epsilon}{_a^b} (\nab{b} u_j) \, \df a \ .
\ee
There is an alternative, particularly useful form of the formula for the coefficients $C_{ij}$. Using the fact that $u_i$ vanishes on $\mathcal{S}_j$ for $j \ne i$, we may write
\be
C_{ij} = -\oint_{\dd\Omega} u_i \, n^a \, \tensor{\epsilon}{_a^b} (\nab{b} u_j) \, \df a \ ,
\ee
so that, using Gauss' theorem and partial integration, we have manifestly symmetric form of the capacitance matrix,
\be\label{eq:symC}
C_{ij} = \epsilon_0 \int_\Omega (\nab{a} u_i) \hat{\epsilon}^{ab} (\nab{b} u_j) \, \vol \ .
\ee

\medskip

An illustrative example of a system for which the capacitance matrix can be exactly calculated is a bispherical capacitor. Suppose that $\Omega$ is a complement of two conducting balls of radii $a_1$ and $a_2$, with centres separated by $b > a_1 + a_2$. Leaving the details of the calculation to the reference \cite{BT} (problems 1.67 and 3.85), we state the final result for the capacitance matrix,
\be\label{eq:bisph}
\mx{C} = \begin{pmatrix} C_{11} & C_{12} \\ C_{21} & C_{22} \end{pmatrix}
\ee
with
\be
\frac{C_{ii}}{4\pi\epsilon_0} = \frac{a_i}{2} + a_i \sinh \xi_i \sum_{\ell = 0}^\infty \exp\Big( -\big( \ell + \frac{1}{2} \big) \, \xi_i \Big) \coth\left( \big( \ell + \frac{1}{2} \big)(\xi_1 + \xi_2) \right)
\ee
and
\be
\frac{C_{12}}{4\pi\epsilon_0} = \frac{C_{21}}{4\pi\epsilon_0} = -a_1 \sinh \xi_1 \sum_{\ell=0}^\infty \frac{\exp\big( -(\ell + \frac{1}{2})(\xi_1 + \xi_2) \big)}{\sinh\big( (\ell + \frac{1}{2})(\xi_1 + \xi_2) \big)} \ ,
\ee
where we have introduced abbreviations $\xi_1$ and $\xi_2$ via
\be
\cosh \xi_1 = \frac{b^2 + a_1^2 - a_2^2}{2 b a_1} \ \quad \textrm{and} \quad \cosh \xi_2 = \frac{b^2 - a_1^2 + a_2^2}{2 b a_2} \ ,
\ee
such that $a_1 \sinh \xi_1 = a_2 \sinh \xi_2$.

\medskip

In our discussion we have left aside some special cases, which may be seen as artificial or pathological configurations from a physical point of view. First, one might wonder what happens with the capacitance matrix when the conductors are sets with empty interior, such as a 2-dimensional plate. The physically reasonable approach is to treat such objects as limiting cases of a regular conductor. For example, Smythe \cite{Smythe} treats disc and line as limits of an ellipsoid. Secondly, there is a class of capacitors with unbounded, but translational invariant conductors. Here one might have a system with one axis of invariance (such as infinite concentric cylinders), in which case the problem is reduced to a 2-dimensional problem, or two axes of invariance (such as a pair of parallel conducting planes), in which case the problem is reduced to a 1-dimensional problem. Since in such cases the integral in (\ref{eq:symC}) will in general diverge, the most sensible solution is to perform the well-defined part of the integration and then interpret the remaining integrand as a ``capacitance density'' (per area or per length).

%%%%%%%%%%%%%%%%%%%%%%%%%%%%%%%%%%%%%%%%%%%%%%%%%%%%%
%%%%%%%%%%%%%%%%%%%%%%%%%%%%%%%%%%%%%%%%%%%%%%%%%%%%%
\section{Structure of the capacitance matrix} %%%%%%%
%%%%%%%%%%%%%%%%%%%%%%%%%%%%%%%%%%%%%%%%%%%%%%%%%%%%%
%%%%%%%%%%%%%%%%%%%%%%%%%%%%%%%%%%%%%%%%%%%%%%%%%%%%%

Once the capacitance matrix has been introduced one can inspect its properties more carefully. Nonnegativity of the diagonal elements,
\be
C_{ii} \ge 0 \ ,
\ee
immediately follows from (\ref{eq:symC}) and the assumption that $\hat{\epsilon}_{ab}$ is positive semi-definite. The functions $u_i$ attain a local minimum on each $\mathcal{S}_j$ for $j \ne i$, $u_i|_{\mathcal{S}_j} = 0$, so that $n^a \nab{a} u_i|_{\mathcal{S}_j} \ge 0$ and, from the assumption (\ref{eps}) about the permittivity tensor, $n^a \tensor{\hat{\epsilon}}{_a^b} \nab{b} u_i \ge n^a \nab{a} u_i$. Hence, (\ref{eq:basicC}) implies that nondiagonal elements,
\be
C_{ij} \le 0
\ee
for each $i \ne j$. Also, it is straightforward to see that the capacitance matrix is block-diagonal.

\smallskip

\btm
The capacitance matrix $\mx{C}$ has a block diagonal form, where each block corresponds to a connected component $\Omega_k \subseteq \Omega$ and the size of the block is equal to the number of connected components of $\dd\Omega_k$. If $\dd\Omega_k$ has only one connected component (physically describing an empty cavity) then the corresponding block is trivial, $1 \times 1$ block $(0)$.
\etm

\noindent
\emph{Proof}. Let $\Omega_k$ and $\Omega_\ell$ be two connected components of $\Omega$, and $\mathcal{S}_i \subseteq \dd\Omega_k$ and $\mathcal{S}_j \subseteq \dd\Omega_\ell$ connected components of their boundaries. Then $u_i|_{\mathcal{S}_j} = 0$ and, vice versa, $u_j|_{\mathcal{S}_i} = 0$. By existence results there is solution for $u_i$ on $\Omega_k$, which we can trivially extend so that $u_i = 0$ on all $\Omega_p$ for $p \ne k$ and, by uniqueness theorems, this is the unique solution (likewise, $u_j$ is zero on all $\Omega_p$ for $p \ne \ell$). Formula (\ref{eq:symC}) then implies that $C_{ij} = 0$. Finally, $\dd\Omega_k$ has only one connected component, the solution is simply $u_k|_{\Omega_p} = \delta_{kp}$, so that $C_{ki} = 0 = C_{ik}$ for all $i \in \{1,\dots,N\}$. \qed

\medskip

One might say that essentially we have $M$ independent problems, with each component $\Omega_i$ described by corresponding block in the capacitance matrix. Now we turn to the properties of sums of elements in rows and elements in columns of the capacitance matrix.

\btm\label{tm:ineqC}
For each connected component $\Omega_i$ of $\Omega$, and connected component $\mathcal{S}_k \subseteq \dd\Omega_i$ we have
\be\label{ineq:Delta}
\Delta_k \defeq C_{kk} - \sum_{\ell = 1,\ne k}^N |C_{k\ell}| \ge 0 \ . %%% \sum_{\substack{\ell=1 \\ \ell\neq k}}^N
\ee
Furthermore, if $\Omega_i$ is bounded then $\Delta_k = 0$ and, conversely, if in addition the permittivity tensor $\hat{\epsilon}_{ab}$ is real analytic and $\Delta_k = 0$ then $\Omega_i$ is necessarily bounded.
\etm 

\noindent
\emph{Proof}. We look at the function $F = \sum_{\ell=1}^N u_\ell$ which satisfies the boundary condition $F|_{\dd\Omega} = 1$. By strong maximum principle $n^a \nab{a} F \le 0$ on $\dd\Omega_i$ and the assumption (\ref{eps}) implies that $n^a \tensor{\epsilon}{_a^b} (\nab{b} F) |_{\mathcal{S}_k} \le 0$. Thus,
\begin{equation*}
\sum_{\ell=1}^N C_{k\ell} = - \oint_{\mathcal{S}_k} n^a \tensor{\epsilon}{_a^b} (\nab{b} F) \, \df a \ge 0 \ ,
\end{equation*}
which can be written in form (\ref{ineq:Delta}) since $C_{kk} \ge 0$ and $C_{k\ell} \le 0$ for all $\ell \ne k$. If $\Omega_i$ is bounded then $F|_{\Omega_i} = 1$ and immediately $\Delta_k = 0$ for each $k$ corresponding to component of boundary $\dd\Omega_i$. Conversely, if $\Delta_k = 0$ on at least one $k$ corresponding to component of boundary $\dd\Omega_i$, then $n^a \tensor{\epsilon}{_a^b} \nab{b} F = 0$ on $\mathcal{S}_k$. This implies that auxiliary function $G = 1 - F$ is zero on $\mathcal{S}_k$ and $n^a \tensor{\epsilon}{_a^b} \nab{b} G = 0$ on $\mathcal{S}_k$, hence by theorem \ref{tm:CauchyEllipt} it follows that $G|_{\Omega_i} = 0$, that is $F|_{\Omega_i} = 1$. This is consistent if $\Omega_i$ is bounded, but in contradiction with the case when $\Omega_i$ is unbounded, as $F(x) \to 0$ as $\norm{x} \to \infty$. \qed

\medskip

An immediate corollary is that for all $i$ and $j$ we have $C_{ii} \ge |C_{ij}|$ and $C_{jj} \ge |C_{ij}|$, implying the inequality
\be\label{ineq:CCC}
C_{ii} C_{jj} \ge C_{ij}^2 \ ,
\ee
which is sometimes derived from the physical assumption about the nonnegativity of the corresponding electrostatic energy.

%%%%%%%%%%%%%%%%%%%%%%%%%%%%%%%%%%%%%%%%%%%%%%%%%%%%%%%%%%
%%%%%%%%%%%%%%%%%%%%%%%%%%%%%%%%%%%%%%%%%%%%%%%%%%%%%%%%%%
\section{Regularization of the capacitance matrix} %%%%%%%
%%%%%%%%%%%%%%%%%%%%%%%%%%%%%%%%%%%%%%%%%%%%%%%%%%%%%%%%%%
%%%%%%%%%%%%%%%%%%%%%%%%%%%%%%%%%%%%%%%%%%%%%%%%%%%%%%%%%% 

\begin{figure}[!ht]
\centering
\begin{tikzpicture}
\draw[lightgray,fill=lightgray] (0,0) circle (2.5);
\draw[very thick,gray,fill=white] (0,0) circle (2.0);
\draw[very thick,gray,fill=lightgray] (1,0) [out=90,in=0] to (0,1.1) [out=180,in=90] to (-1.3,0) [out=-90,in=180] to (0,-0.9) [out=0,in=-90] to (1,0); 

\node at (7,0) {$\mx{C} = \begin{pmatrix} C_{11} & -C_{11} \\ -C_{11} & C_{11} \end{pmatrix}$};
\end{tikzpicture}
\caption{Left: Schematic representation (cross-section) a simple capacitor with $(M,N,P) = (1,2,2)$. Right: The corresponding capacitance matrix.}
\label{fig:cap}
\end{figure}
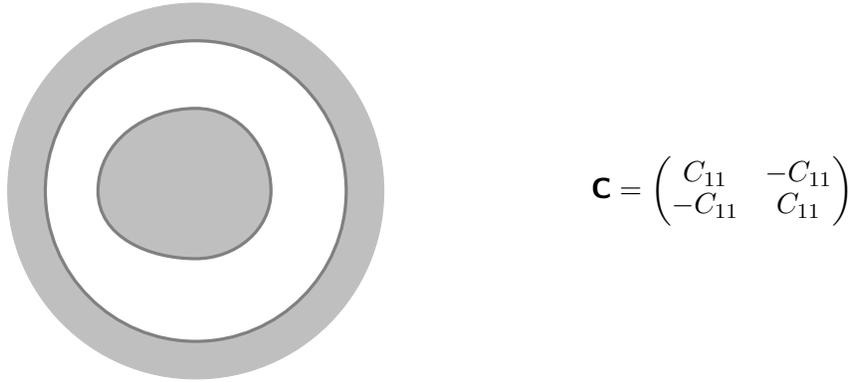

Some problems demand inverse of the capacitance matrix, given that we start with charges on conductors and try to relate them to potential differences. A technical obstacle to this task, however, is that the capacitance matrix (\ref{eq:basicC}) is in general not regular. This was already revealed by the theorem \ref{tm:ineqC}, since for any surface $\mathcal{S}_k$ in the boundary of a bounded component $\Omega_i$ the sum of the $k$-th row, as well as the sum of the $k$-th column in $\mx{C}$, are zero. In other words, regularity of the capacitance matrix is spoiled by the presence of cavities. An elementary example is a capacitor with two compact conducting surfaces, one inner and one outer, as shown in Figure \ref{fig:cap}. The corresponding capacitance matrix, according to theorem \ref{tm:ineqC}, is singular as its rank is 1 and determinant is equal to 0.

\smallskip

\smallskip

The results derived in the previous section are still not enough for the systematic treatment of the matrix regularization. In order to infer sharper inequalities we must impose a stronger assumption about the permittivity tensor.

\smallskip

\btm\label{tm:nonzeroC}
Suppose that the permittivity tensor $\hat{\epsilon}_{ab}$ is real analytic and let $\Omega_i$ be a connected (either bounded or external) component of $\Omega$. If the boundary $\dd\Omega_i$ consists of at least two connected components, then all elements of the corresponding block in the capacitance matrix are strictly non-zero.
\etm

\noindent
\emph{Proof}. First we look at nondiagonal elements $C_{k\ell}$, where $k$ and $\ell$ indices correspond to components of the boundary $\dd\Omega_i$, defined by the integral (\ref{eq:basicC}) of normal derivatives $n^a \tensor{\hat{\epsilon}}{_a^b} \nab{b} u_\ell$ over the surface $\mathcal{S}_k$. As on this surface the function $u_\ell$ attains a local minimum, we know that $n^a \tensor{\hat{\epsilon}}{_a^b} \nab{b} u_\ell \ge 0$ over the whole $\mathcal{S}_k$. Hence $C_{k\ell} = 0$ holds if and only if $n^a \tensor{\hat{\epsilon}}{_a^b} \nab{b} u_\ell = 0$ over the whole $\mathcal{S}_k$. But, if the latter holds, the fact that $u_\ell|_{\mathcal{S}_k} = 0$, together with the theorem \ref{tm:CauchyEllipt}, implies $u_\ell = 0$ over the whole $\Omega_i$, in contradiction with the rest of the boundary conditions, namely $u_\ell|_{\mathcal{S}_\ell} = 0$. For diagonal elements we look at the auxiliary function $v_k \defeq 1 - u_k$, which is again a solution of (\ref{eq:hardie}), satisfies boundary conditions $v_k|_{\mathcal{S}_\ell} = 1 - \delta_{k\ell}$, and has local minimum on $\mathcal{S}_k$. Hence, according to the same argument as with the nondiagonal elements we have $n^a \nab{a} v_k > 0$, that is $n^a \nab{a} u_k < 0$ on $\mathcal{S}_k$, whence $C_{kk} > 0$. \qed

\smallskip

Let us look more closely at one square block of the capacitance matrix, say $\mx{A}$, of size $r \times r$ (without loss of generality, we shall assume that $\mx{A}$ is the first block in $\mx{C}$), corresponding to a connected component $\Omega_s$. For any $\mx{x} \in \rr^r$ we have
\bal
\mx{x}\trans\mx{A}\mx{x} & = \sum_{k=1}^r x_k^2 \left( \Delta_k + \sum_{\ell\neq k}^r |C_{k\ell}| \right) - \sum_{i=1}^r \sum_{j\neq i}^r x_i x_j |C_{ij}| \nonumber\\
 & = \sum_{k=1}^r \Delta_k x_k^2 + \frac{1}{2} \, \sum_{i=1}^r \sum_{j \ne i}^r |C_{ij}| (x_i - x_j)^2 \ge 0 \ .
\eal
If $\Omega_s$ is bounded then $\Delta_k = 0$ for all $k \in \{1,\dots,r\}$, and if in addition $\Omega_s$ has two or more connected components of boundary $\dd\Omega_s$ then, according to the theorem \ref{tm:nonzeroC}, all $C_{ij} \ne 0$. Thus, under all these assumption, $\mx{x}\trans\mx{A}\mx{x} = 0$ if and only if $x_i = x_j$ for all pairs $i,j \in \{1,\dots,r\}$, or
\be
\ker\mx{A} = \left\{ (\lambda,\dots,\lambda) \in \rr^r \mid \lambda \in \rr \right\} \ .
\ee
Note that this result is a direct descendant of theorem \ref{tm:CauchyEllipt}. Furthermore, the matrix $\mx{A}'$, obtained by removal of one row and one column from $\mx{A}$, is a regular matrix. In order to prove this, we may start from the decomposition
\be
\mx{x}\trans\mx{A}\mx{x} = {\mx{x}'}\trans\mx{A}'\mx{x}' + \left( A_{rr} x_r + 2 \sum_{i=1}^{r-1} x_i A_{ir} \right) x_r \ .
\ee
Now, suppose that there is $\mx{x}' \in \rr^{r-1}$, such that ${\mx{x}'}\trans\mx{A}'\mx{x}' = 0$. Here one can simply choose $x_r = 0$, so that $\mx{x} \in \ker\mx{A}$. But, according to result above, this is possible if and only if $\lambda = 0$, that is $\mx{x} = \0$. Thus $\ker\mx{A}'$ is trivial.

\smallskip

Finally, it is not too difficult to prove that the block $\mx{C}_\mathrm{e}$ of the capacitance matrix $\mx{C}$, corresponding to the unbounded component $\Omega_\mathrm{e}$, is necessarily regular. Suppose that $\mx{C}_\mathrm{e}\,\mx{x} = \0$. We know that a trivial potential, $\mx{x} = \0$, solves the electrostatic problem in which all the connected components of the boundary of the exterior region are electrically neutral (and the potential is assumed to decay at infinity), and by uniqueness theorems this is also the only one. Hence $\ker\mx{C}_\mathrm{e}$ is trivial and $\mx{C}_\mathrm{e}$ is regular.

\smallskip

Before we proceed we shall introduce a slight modification of the notation, which should allow easier following of the discussion. Let $M'$ be a number of bounded connected components of $\Omega$. In another words, $M' = M$ if there is no exterior region $\Omega_{\mathrm{e}}$ and $M' = M-1$ if there is $\Omega_{\mathrm{e}} \ne \emptyset$. Furthermore, we denote by $r_a$ the number of connected components of $\dd\Omega_a$ (where $\Omega_a$ is bounded) for each $a \in \{1,\dots,M'\}$ and by $f$ the number of connected components of $\dd\Omega_{\mathrm{e}}$, so that
\be
f + \sum_{a=1}^{M'} r_a = N \ .
\ee
Connected components of $\dd\Omega_a$ may be denoted by $\{\mathcal{S}_1^{(a)},\dots,\mathcal{S}_{r_a}^{(a)}\}$ and pertaining potentials by $\{\varphi_1^{(a)},\dots,\varphi_{r_a}^{(a)}\}$, where we follow a convention that the last one $\mathcal{S}_{r_a}^{(a)}$, in case of bounded $\Omega_a$, is a bounding, outermost surface. Accordingly, as the capacitance matrix has block-diagonal form, we may consistently use the notation $C^{(a)}_{ij}$, where the superscript denotes the domain $\Omega_a$, that is a corresponding block.

\smallskip

\emph{Regularization, version 1}. The simplest route to a regularized capacitance matrix relies on the observations made in the first part of this section. Note that for all $a \in \{1,\dots,M'\}$ and $i \in \{1,\dots,r_a\}$ we have
\be
Q_i^{(a)} = \sum_{j=1}^{r_a - 1} C^{(a)}_{ij} (\varphi_j^{(a)} - \varphi_{r_a}^{(a)}) \ ,
\ee
and for all $i \in \{1,\dots,f\}$
\be
Q_i^{(\mathrm{e})} = \sum_{j=1}^f C^{(\mathrm{e})}_{ij} (\varphi_j^{(\mathrm{e})} - \varphi_\infty) \ .
\ee
Thus, if we remove a row and a column passing through a $r_a$-th row and $r_a$-th column in each $r_a \times r_a$ block, corresponding to a bounded $\Omega_a$, of the matrix $\mx{C}$, the result will be a $(N - M') \times (N - M')$ regular matrix $\mx{\widehat{C}}$. Simultaneously, this means that each potential difference $\varphi_k^{(a)} - \varphi_\infty$ on surface $\mathcal{S}_k^{(a)}$, part of the boundary $\dd\Omega_a$ which is not its bounding, outermost surface, must be replaced with the potential difference $\varphi_k^{(a)} - \varphi_{r_a}^{(a)}$. One drawback of the matrix $\mx{\widehat{C}}$ is that the charges $Q_{r_a}^{(a)}$ on bounding surfaces are not directly produced by the product $\mx{\widehat{C}}\bm{\varphi}$, but must be calculated separately via Gauss' law in integral form, that is
\be
Q_{r_a}^{(a)} = -\sum_{i=1}^{r_a - 1} Q_i^{(a)} \ .
\ee
Another source of discontent with the matrix $\mx{\widehat{C}}$ is that product $\mx{\widehat{C}}\bm{\varphi}$ does not result in a sum of charges on different boundaries of the same connected conductor $K_p$. This motivates another version of the regularization.

\smallskip

\emph{Regularization, version 2}. Each bounded conductor $K_p$ has an ``outer'' part of the boundary, which it shares with the surrounding part of the domain $\Omega$, and $i_p \ge 0$ ``inner'' parts of the boundary, bounding cavities $\{\Omega_{\kappa(p,1)},\dots,\Omega_{\kappa(p,i_p)}\}$ inside of it (if there are any). In order to obtain the total charge $Q(K_p)$ on $K_p$ we need to sum the charge on the outer boundary with the charges on each of $i_p$ inner boundaries of the conductor $K_p$. This translates to the following recipe for the alteration of the matrix $\mx{\widehat{C}}$: Subtract from a row corresponding to the outer boundary of $K_p$ each row corresponding to a non-outermost connected component of $\dd\Omega_{\kappa(p,j)}$ (here we are relying on the fact that the charge on the outermost boundary of the cavity is minus sum of charges on all inner boundaries of that cavity). The procedure must be performed from the innermost conductors in the system, up to the outermost. The resulting matrix, which we shall denote by $\mx{\widetilde{C}}$, will have unaltered determinant, $\det(\mx{\widetilde{C}}) = \det(\mx{\widehat{C}})$, thus it will remain regular. Note that the matrix $\mx{\widetilde{C}}$ is, in general, not diagonal.

\smallskip

\emph{Regularization, version 3}. Still, one might look for another regularized matrix $\mx{\check{C}}$ which combines traits of all the previous matrices, namely the one that remains symmetric (just as the matrix $\mx{\widehat{C}}$) and produces the total charges on connected conductors (just as the matrix $\mx{\widetilde{C}}$) by acting on potential differences $\varphi_i - \varphi_\infty$ (just as the original matrix $\mx{C}$). Such matrix is, indeed, possible to construct, starting from the matrix $\mx{\widetilde{C}}$ with the following recipe: Subtract from a row corresponding to the outer boundary of $K_p$ each column corresponding to a non-outermost connected component of $\dd\Omega_{\kappa(p,j)}$. Essentially, we are making here rearrangement of the matrix, corresponding to the potential decomposition $\varphi_i - \varphi_{r_a} = (\varphi_i - \varphi_\infty) - (\varphi_{r_a} - \varphi_\infty)$. As the procedure treats rows and columns of the capacitance matrix $\mx{C}$ in a symmetric fashion, the result is again a symmetric matrix\footnote{Note that the ``coefficients of capacitance'' $c_{ij}$ from \cite{BT}, equation (3.b.3), are in fact elements of the matrix $\mx{\check{C}}$.}. Furthermore, the procedure preserves the value of the determinant, so that we have $\det(\mx{\check{C}}) = \det(\mx{\widetilde{C}}) = \det(\mx{\widehat{C}})$.

\smallskip

\begin{figure}[!ht]
\centering
\begin{tikzpicture}[scale=0.8]
\draw[thick,gray,fill=lightgray] (0,0) ellipse (4 and 3);
\draw[thick,gray,fill=white] (-2,0) circle (1.3);
\draw[thick,gray,fill=white] (2,0) ellipse (1.3 and 1.5);
\draw[thick,gray,fill=lightgray] (-2,0) ellipse (0.7 and 0.5);
\draw[thick,gray,fill=lightgray] (2,0) circle (0.7);
\node at (0,2) {\small $K_3$};
\node at (-2,0) {\small $K_1$};
\node at (2,0) {\small $K_2$};
\node at (-2.9,2.5) {\small $\mathcal{S}_5$};
\node at (-1.75,0.7) {\small $\mathcal{S}_1$};
\node at (2.25,0.9) {\small $\mathcal{S}_3$};
\node at (-0.95,-1.25) {\small $\mathcal{S}_2$};
\node at (1.25,-1.5) {\small $\mathcal{S}_4$};
\end{tikzpicture}
\caption{Schematic cross-section of a system of conductors (conductor with two cavities and in each of them one additional conductor).}
\label{fig:example}
\end{figure}
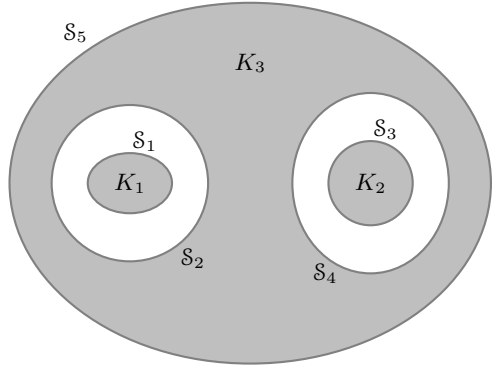

All three versions of a regularized capacitance matrix may be illustrated with a simple example, sketched in Figure \ref{fig:example}.

\medskip

\noindent
The initial capacitance matrix is (for clarity we omit zeros in the matrices)
\be
\mx{C} = \begin{pmatrix} C_{11} & -C_{11} & & & \\ -C_{11} & C_{11} & & & \\ & & C_{33} & -C_{33} & \\ & & -C_{33} & C_{33} & \\ & & & & C_{55} \end{pmatrix}
\ee
and the corresponding regularized capacitance matrices are 
\be
\mx{\widehat{C}} = \begin{pmatrix} C_{11} & & \\ & C_{33} & \\ & & C_{55} \end{pmatrix} \qqd \begin{pmatrix} Q_1 \\ Q_3 \\ Q_5 \end{pmatrix} = \mx{\widehat{C}} \begin{pmatrix} \varphi_1 - \varphi_5 \\ \varphi_3 - \varphi_5 \\ \varphi_5 - \varphi_\infty \end{pmatrix} \ ,
\ee
\be
\mx{\widetilde{C}} = \begin{pmatrix} C_{11} & & \\ & C_{33} & \\ -C_{11} & -C_{33} & C_{55} \end{pmatrix} \qqd \begin{pmatrix} Q(K_1) \\ Q(K_2) \\ Q(K_3) \end{pmatrix} = \mx{\widetilde{C}} \begin{pmatrix} \varphi_1 - \varphi_5 \\ \varphi_3 - \varphi_5 \\ \varphi_5 - \varphi_\infty \end{pmatrix} \ ,
\ee
and
\be
\mx{\check{C}} = \begin{pmatrix} C_{11} & & -C_{11} \\ & C_{33} & -C_{33} \\ -C_{11} & -C_{33} & C_{11} + C_{33} + C_{55} \end{pmatrix} \qqd \begin{pmatrix} Q(K_1) \\ Q(K_2) \\ Q(K_3) \end{pmatrix} = \mx{\check{C}} \begin{pmatrix} \varphi_1 - \varphi_\infty \\ \varphi_3 - \varphi_\infty \\ \varphi_5 - \varphi_\infty \end{pmatrix} \ .
\ee
Here we have $Q(K_1) = Q_1$, $Q(K_2) = Q_3$ and $Q(K_3) = Q_2 + Q_4 + Q_5 = -Q_1 - Q_3 + Q_5$.

\medskip

%%%%%%%%%%%%%%%%%%%%%%%%%%%%%%%%%%%%%%%%%%%%%%%%%%%%%%%%
%%%%%%%%%%%%%%%%%%%%%%%%%%%%%%%%%%%%%%%%%%%%%%%%%%%%%%%%
\section{Charge exchange between the conductors} %%%%%%%
%%%%%%%%%%%%%%%%%%%%%%%%%%%%%%%%%%%%%%%%%%%%%%%%%%%%%%%%
%%%%%%%%%%%%%%%%%%%%%%%%%%%%%%%%%%%%%%%%%%%%%%%%%%%%%%%%

Capacitors serve to separate charges and the elements of the capacitance matrix should in some sense quantify how much is a given system of conductors successful in this respect. A direct insight is given by a simple procedure: One connects a pair of conductors with a thin conducting wire and measures the charge $\Delta Q$ exchanged through the wire. The assumption here is that the wire used in the experiment should be ``thin'' in a sense that it only serves as bridge for the exchange of the charge and that it negligibly perturbs the capacitance matrix during the experiment. Suppose that the initial charges on the pair of conductors were $Q_a$ and $Q_b$, while their potentials were $\varphi_a$ and $\varphi_b$. We claim that $\Delta Q$ is given by
\be\label{eq:DeltaQ}
\Delta Q = \frac{\Delta\varphi}{\check{C}_{aa}^{-1} + \check{C}_{bb}^{-1} - 2\check{C}_{ab}^{-1}}
\ee
with $\Delta Q = Q_a - Q'_a$ and $\Delta\varphi = \varphi_a - \varphi_b$. The attack on the problem relies on the properties of the matrix $\mx{\check{C}}$, which allows us to start from the potentials expressed with charges. Prior to connecting we have
\bal
\varphi_a - \varphi_\infty & = \check{C}_{aa}^{-1} Q_a + \check{C}_{ab}^{-1} Q_b + \sum_{i \ne a,b} \check{C}_{ai}^{-1} Q_i \ , \\
\varphi_b - \varphi_\infty & = \check{C}_{ba}^{-1} Q_a + \check{C}_{bb}^{-1} Q_b + \sum_{i \ne a,b} \check{C}_{bi}^{-1} Q_i \ ,
\eal
and after the connecting we must make replacements $\varphi_a \to \varphi'_a$, $\varphi_b \to \varphi'_b$, $Q_a \to Q'_a$ and $Q_b \to Q'_b$. Hence
\bal
\varphi_a - \varphi'_a & = \check{C}_{aa}^{-1} (Q_a - Q'_a) + \check{C}_{ab}^{-1} (Q_b - Q'_b) \ , \label{eq:delphi1}\\
\varphi_b - \varphi'_b & = \check{C}_{ba}^{-1} (Q_a - Q'_a) + \check{C}_{bb}^{-1} (Q_b - Q'_b) \ . \label{eq:delphi2}
\eal
As $\mx{\check{C}}^{-1}$ is symmetric, $\varphi'_a = \varphi'_b$ and $Q_a + Q_b = Q'_a + Q'_b$, that is $Q_a - Q'_a = Q'_b - Q_b$, equation (\ref{eq:DeltaQ}) follows from subtraction of equations (\ref{eq:delphi1}) and (\ref{eq:delphi2}).

\smallskip

As a special, more concrete example, we may take a system of two compact disjoint conductors in a vacuum, each with one connected component of the boundary (that is, without cavities). Here we have
\be
\mx{\check{C}} = \mx{C} = \begin{pmatrix} C_{aa} & C_{ab} \\ C_{ab} & C_{bb} \end{pmatrix} \ ,
\ee
so that
\be
\Delta Q = \frac{C_{aa}C_{bb} - C_{ab}^2}{C_{aa} + C_{bb} + 2C_{ab}}\,\Delta\varphi \ ,
\ee
in agreement with the result presented in \cite{LLED}.

%%%%%%%%%%%%%%%%%%%%%%%%%%%%%%%%%%%%%%%%%%%%%%%%%%%%%
%%%%%%%%%%%%%%%%%%%%%%%%%%%%%%%%%%%%%%%%%%%%%%%%%%%%%
\section{Real-life approximations} %%%%%%%
%%%%%%%%%%%%%%%%%%%%%%%%%%%%%%%%%%%%%%%%%%%%%%%%%%%%%
%%%%%%%%%%%%%%%%%%%%%%%%%%%%%%%%%%%%%%%%%%%%%%%%%%%%%

Calculation of the elements of the capacitance matrix is, except in some special cases, technically highly involved. Bispherical capacitor, featuring subcases of disjoint conducting balls and a conducting ball in a spherical cavity with conducting walls, belongs to a rare class which allows exact solution written in an analytic form. However, difficulties arise even at examples with deceiving geometric simplicity, such as a capacitor consisting of two parallel coaxial discs. History of the latter problem goes back to Maxwell \cite{Maxwell}, over Love, who managed to reduce the corresponding Laplace problem to a solution of a linear integral equation \cite{Love49}, to various modern treatments \cite{Hutson63,Rao05,PCDiLM16,Paffuti17}. Another source of complications arises from the permittivity tensor $\epsilon_{ab}$. Suppose, for example, that a spherical capacitor is filled with anisotropic homogeneous dielectric, such that in a Cartesian coordinate system $\epsilon_{\mu\nu} = \mathrm{diag}\,(\epsilon_1,\epsilon_2,\epsilon_3)$. In such case a simple coordinate rescaling $x'^i = x^i/\sqrt{\epsilon_i}$ reduces the differential equation (\ref{eq:hardie}) to the basic Laplace equation, but spherical boundaries become ellipsoids of the form $r(\mathbf{\hat{n}}) = a (\hat{\epsilon}_{ij} \hat{n}^i \hat{n}^j)^{-1/2}$. Analytic approaches to such problems range from perturbative techniques \cite{Erma63} to geometric inequalities \cite{PS}.

\medskip

Let us turn to class of capacitors consisting of two closely separated pair of finite conducting plates. A pragmatic approach, often used to bypass calculational difficulties, is to neglect fringing fields (see e.g.~discussion in \cite{SBS86}). In this approximation the corresponding capacitance matrix is, up to the order $O(\delta)$ of the ``regulator'' parameter $\delta > 0$, given by
\be
\mx{C} \approx \begin{pmatrix} C_{11} & -(C_{11} - \delta) \\ -(C_{11} - \delta) & C_{11} \end{pmatrix} \ .
\ee
Intuitively, $\delta = 0$ corresponds to the capacitance matrix of enclosed system of two conductors, while $\delta > 0$ measures the amount of the charge that has distributed on the ``outer'' sides of the capacitor. A paradigmatic example is a planar capacitor, consisting of two parallel plates, each of area $A$, separated by $h$, with space between them filled with dielectric described by the permittivity tensor $\epsilon_{ab}$. We orient the Cartesian coordinate system so that $z$-axis is perpendicular to the plates. Now, assuming that the charge on the ``outer'' side of a capacitor plate is negligible, we have
\be
Q_1 \approx -\int \epsilon^{ab} n_a \nab{b} \Phi |_{z=0} \, \df a = -\int \epsilon^{zi}(x,y,0) \, \dd_i \Phi(x,y,0) \, \df a
\ee
and
\be
\varphi_1 - \varphi_2 = -\int_0^h \dd_z \Phi \, \df z \ .
\ee
This allows us to write
\be
\frac{C_{11}}{C_0} \approx \frac{\displaystyle{\frac{1}{A} \, \int \hat{\epsilon}^{zi}(x,y,0) \, \dd_i \Phi(x,y,0) \, \df a}}{\displaystyle{\frac{1}{h} \, \int_0^h \dd_z \Phi \, \df z}} \ ,
\ee
where we have introduced the abbreviation $C_0 = \epsilon_0 A/h$, approximate capacitance of the vacuum plate capacitor. We may single out several special subcases:

\begin{itemize}
\item[(a)] isotropic dielectric, homogeneous in $z$-direction: equation (\ref{eq:hardie}) admits a simple solution of the form $\Phi(z) = \alpha z + \beta$, so that
\be
\frac{C_{11}}{C_0} \approx \left< \hat{\epsilon} \right> \defeq \frac{1}{A} \, \iint \df x\,\df y\, \hat{\epsilon}(x,y)
\ee

\item[(b)] isotropic dielectric with $\hat{\epsilon} = \hat{\epsilon}(z)$: equation (\ref{eq:hardie}) admits a solution of the form
\be
\Phi(z) = \varphi_1 + \int_0^z \df z' \, \frac{K}{\hat{\epsilon}(z')} \ ,
\ee
where the constant $K$ is fixed by the condition $\Phi(h) = \varphi_2$, so that
\be
\frac{C_{11}}{C_0} \approx \frac{1}{\left< 1/\hat{\epsilon} \right>} \defeq \frac{1}{\displaystyle{\frac{1}{h} \, \int_0^h \frac{1}{\hat{\epsilon}(z)} \, \df z}}
\ee

\item[(c)] homogeneous, but not necessarily isotropic dielectric: again, equation (\ref{eq:hardie}) admits a simple solution of the form $\Phi(z) = \alpha z + \beta$, so that
\be
\frac{C_{11}}{C_0} \approx \hat{\epsilon}_{zz} \ .
\ee
\end{itemize}

\medskip

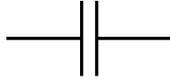
\begin{figure}[!ht]
\centering
\begin{tikzpicture}
\draw[very thick] (0,0) -- (1,0);
\draw[very thick] (1,-0.5) -- (1,0.5);
\draw[very thick] (1.2,-0.5) -- (1.2,0.5);
\draw[very thick] (1.2,0) -- (2.2,0);
\end{tikzpicture}
\caption{Capacitor symbol.}
\label{fig:capsym}
\end{figure}

Another challenge is to describe a system of capacitors in the lumped-element limit, appearing as a part of an electric circuit. Each capacitor (pair of conducting plates) is represented by the symbol depicted in Figure \ref{fig:capsym}.

\noindent
Here we assume that all capacitors are well separated, such that the capacitance matrix may be approximated by a block-diagonal matrix (with each block corresponding to each capacitor). An insight into such an approximation is provided by the bispherical capacitor, mentioned in section 4. Namely, expanding (\ref{eq:bisph}) with respect to the parameter $\eta \defeq \sqrt{a_1 a_2}/b$ up to order $O(\eta^2)$, we get
\be
\mx{C} \approx 4\pi \epsilon_0
\begin{pmatrix}
a_1 \left(1 + \eta^2 \right) & - \eta \sqrt{a_1 a_2} \\
- \eta \sqrt{a_1 a_2} & a_2 \left(1 + \eta^2 \right) \ .
\end{pmatrix}
\ee
We see that in the limit when $\eta \to 0$, the capacitance matrix is of block-diagonal form with leading correction term in the off-diagonal elements. Any two elements of an electric circuit may be combined by two elementary types of connection, series and parallel. We shall investigate such combinations, by calculating the effective capacitance defined as a ratio $\Delta Q/\Delta \varphi$, that was introduced and analyzed in the previous section.

\medskip

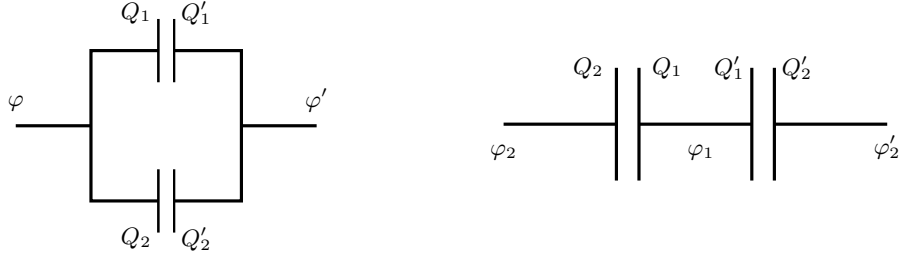
\begin{figure}[!ht]
\centering

\begin{tabular}{ccc}
\begin{tikzpicture}
\draw[very thick] (0,0) -- (4,0);
\draw[very thick,fill=white] (1,-1) rectangle (3,1);
\draw[very thick,fill=white] (1.9,0.6) rectangle (2.1,1.4);
\draw[very thick,fill=white] (1.9,-0.6) rectangle (2.1,-1.4);
\draw[white,fill=white] (1.92,-1.5) rectangle (2.08,1.5);
\node at (0,0.3) {\small $\varphi$};
\node at (4,0.35) {\small $\varphi'$};
\node at (1.6,1.5) {\small $Q_1$};
\node at (2.4,1.5) {\small $Q'_1$};
\node at (1.6,-1.5) {\small $Q_2$};
\node at (2.4,-1.5) {\small $Q'_2$};
\end{tikzpicture}

& \hspace{30pt} &

\begin{tikzpicture}[scale=1.5]
\draw[very thick] (0,0) -- (1,0);
\draw[very thick] (1,-0.5) -- (1,0.5);
\draw[very thick] (1.2,-0.5) -- (1.2,0.5);
\draw[very thick] (1.2,0) -- (2.2,0);
\draw[very thick] (2.2,-0.5) -- (2.2,0.5);
\draw[very thick] (2.4,-0.5) -- (2.4,0.5);
\draw[very thick] (2.4,0) -- (3.4,0);
\node at (0,-0.25) {\small $\varphi_2$};
\node at (1.75,-0.25) {\small $\varphi_1$};
\node at (3.4,-0.2) {\small $\varphi'_2$};
\node at (0.75,0.5) {\small $Q_2$};
\node at (1.45,0.5) {\small $Q_1$};
\node at (2,0.5) {\small $Q'_1$};
\node at (2.6,0.5) {\small $Q'_2$};
\draw[white] (0,-1.1) circle (0.1);
\end{tikzpicture}
\end{tabular}

\caption{Left: Pair of capacitors, connected in parallel. Right: Pair of capacitors, connected in series.}
\label{fig:parser}
\end{figure}

\medskip

\emph{Parallel} (Figure \ref{fig:parser}, left). Essentially, this is again a two-piece capacitor. The total charges on each of the joined plates, under the approximation explained above, are
\be
\begin{pmatrix} Q_1 + Q_2 \\ Q'_1 + Q'_2 \end{pmatrix} = \begin{pmatrix} C_1 + C_2 & -(C_1 + C_2 - 2\delta) \\ -(C_1 + C_2 - 2\delta) & C_1 + C_2 \end{pmatrix} \begin{pmatrix} \varphi \\ \varphi' \end{pmatrix}
\ee
If one connects the left and right side of circuit, the exchanged charge divided by the potential difference (\ref{eq:DeltaQ}), is
\be
\frac{\Delta Q}{\Delta\varphi} = \frac{(C_1+C_2)^2 - (C_1 + C_2 - 2\delta)^2}{4\delta} = C_1 + C_2 - \delta
\ee
and in the limit when $\delta \to 0$ we have 
\be
\lim_{\delta\to 0} \, \frac{\Delta Q}{\Delta\varphi} = C_1 + C_2 \ .
\ee
This is the well-know formula for the total capacitance of the capacitors connected in parallel. 

\medskip

\emph{Series} (Figure \ref{fig:parser}, right). In this case we end with a system of three conductors (see discussion in \cite{JS15}). The total charges on each of the conductors are given by
\be
\begin{pmatrix} Q_1 + Q'_1 \\ Q_2 \\ Q'_2 \end{pmatrix} = \begin{pmatrix} C_1 + C_2 & -(C_1 - \delta) & -(C_2 - \delta) \\ -(C_1 - \delta) & C_1 & 0 \\ -(C_2 - \delta) & 0 & C_2 \end{pmatrix} \begin{pmatrix} \varphi_1 \\ \varphi_2 \\ \varphi'_2 \end{pmatrix} \ .
\ee
Now, using the same procedure as above, if one connects the left and right side of circuit, the exchanged charge divided by the potential difference, in the limit when $\delta \to 0$, is
\be
\lim_{\delta\to 0} \, \frac{\Delta Q}{\Delta\varphi} = \frac{1}{\displaystyle{\frac{1}{C_1} + \frac{1}{C_2}}} \ .
\ee
This is the well-know formula for the total capacitance of the capacitors connected in series.

%%%%%%%%%%%%%%%%%%%%%%%%%%%%
%%%%%%%%%%%%%%%%%%%%%%%%%%%%
\section{Conclusion} %%%%%%%
%%%%%%%%%%%%%%%%%%%%%%%%%%%%
%%%%%%%%%%%%%%%%%%%%%%%%%%%%

In this paper we have presented a thorough, rigorous analysis of the capacitance matrix and its use for practical problems. Although some properties, such as the symmetry of the matrix $\mx{C}$ and corresponding non-strict inequalities, are classical results (some going back to Maxwell \cite{Maxwell}), here we have pushed these conclusions one step further. In order to achieve this, we have unearthed many hidden assumptions, both geometric related to properties of the domain $\Omega$ and analytic related to the properties of the permittivity tensor $\epsilon_{ab}$. This has allowed us to prove strict inequalities, identify kernel of the capacitance matrix and propose several procedures for its regularization. Utilization of the regularized capacitance matrix is illustrated in section 7, with formula for the charge exchanged between the conductors. Furthermore, we have discussed the approximation schemes, featuring lumped element limit and neglect of the fringing fields, which admit one to pass from capacitance matrix to standard rules for capacitors in electric circuits. As the electric circuit may contain some other elements, such as resistors and inductors, the full treatment usually demands introduction of the corresponding conductance and inductance matrices, sharing some similar properties with the capacitance matrix.

\smallskip

Apart from the simplest, electrostatic problems, the capacitance matrix appears in various other branches of classical electrodynamics. For example, Maxwell's equations applied to the TEM modes in waveguides \cite{Jack} are reduced to the Laplace equation, and the capacitance matrix again naturally appears in the equation governing the dynamics of the system \cite{Olyslager}. Looking from even broader perspective, any physical problem which is defined with solutions of the Laplace equation admits introduction of some analogue of the capacitance matrix. The main problem is that such matrix does not have to contain any physically useful information. For example, one might look at the stationary solutions of the heat equation, on a domain whose boundaries is kept on fixed temperatures. Then the total heat flux (``charge'') on each component of the boundary will be related to the boundary temperatures via capacitance matrix. We note in passing that there is so-called ``thermal capacitance matrix'' \cite{BhMu,coffee}, unrelated to the geometric matrix defined here. 

\smallskip

Our hope is that this work might stimulate further research on the properties of the capacitance matrix and encourage its use in the undergraduate courses on classical electrodynamics.

\ack
We would like to thank Edgardo Franzin for thorough reading of the first draft and many comments which helped us to improve the paper. Also, we would like to thank Marko Erceg for several useful remarks. The work of one author (B.K.) was supported by the Croatian Science Foundation Grant No. IP-2016-06-5885 SynthMagIA.

%%%%%%%%%%%%%%%%%%%%%%%%%%%%%%%%%%%%%%%%%%%%%%%%%%%%%%%%%%
%%%%%%%%%%%%%%%%%%%%%%%%%%%%%%%%%%%%%%%%%%%%%%%%%%%%%%%%%%

\renewcommand{\thesection}{Appendix \Alph{section}} %%% correction of the Appendix section format 
%%%%%%%%%%%%%%%%%%%%%%%%%%%%%%%%%%%%%%%%%%%%%%%%%%%%%%%%%%%%%%%%%
%%%%%%%%%%%%%%%%%%%%%%%%%%%%%%%%%%%%%%%%%%%%%%%%%%%%%%%%%%%%%%%%%
\appendixx{Existence, uniqueness, triviality} %%%%%%%%%%%%%%%%%%%%%
%%%%%%%%%%%%%%%%%%%%%%%%%%%%%%%%%%%%%%%%%%%%%%%%%%%%%%%%%%%%%%%%%
%%%%%%%%%%%%%%%%%%%%%%%%%%%%%%%%%%%%%%%%%%%%%%%%%%%%%%%%%%%%%%%%%
\renewcommand{\thesection}{\Alph{section}}  %%% for proper alphabetic "numbering" of theorems and lemmas 

Analysis of electrostatic potentials is grounded in the theory of harmonic functions and their generalizations. Let us briefly review the Dirichlet problem for a class of partial differential equations relevant for this paper.  We assume that domain is a nonempty open set $\Omega \subseteq \rr^3$, such that either $\Omega$ or its complement $\rr^3 - \Omega$ is bounded (the latter is known as the \emph{exterior problem}). Here we look for a solution of an elliptic partial differential equation
\be\label{eq:ellip}
\nab{a} (\hat{\epsilon}^{ab} \nabla_{b} u) = 0 \ \, \textrm{on} \ \Omega \ ,
\ee
a function $u \in C^2(\Omega) \cap C^0(\cl{\Omega})$ (that is, twice differentiable in the interior of the domain and continuous up to the boundary $\dd\Omega$), which satisfies the boundary condition
\be
u = g \ \, \textrm{on} \ \dd\Omega
\ee
for some $g \in C^0(\dd\Omega)$ and, if $\Omega$ is unbounded, which converges to some fixed constant as $\norm{x} \to \infty$. Obviously, the problem naturally splits into $M$ independent Dirichlet problems, one for each connected component of the domain $\Omega$. Now, we are faced with two central questions: Does such solution exist and, if it does, is it unique?

\smallskip

\emph{Existence}. Given that throughout the paper we assume that tensor $\hat{\epsilon}_{ab}$ is at least of class $C^1$, the main source of concern here is (geometric) smoothness of the boundary $\dd\Omega$. A simple example of a Dirichlet problem without solution was pointed out already by Zaremba \cite{Zaremba11} in 1911, the one in which $\Omega$ is an open ball $B$ with its center removed, and where $g$ is equal to $0$ on $\dd B$ and equal to $1$ in its center. A more subtle example, with connected boundary $\dd\Omega$, was constructed by Lebesgue \cite{Lebesgue13}, now usually referred to as a Lebesgue spine (see Remark 6.6.17.~in \cite{AG}). One might argue that, for example, a domain with Lebesgue spine is a mere mathematical pathology, unattainable to physical experiments. Note, however, that in our models we often use domains with edges, corners and spikes (think of a simple cube), so that any theory restricted only to smooth boundaries would be too narrow in its scope. Unfortunately, textbooks on classical electrodynamics are usually silent with respect to problem of existence of solutions and all related issues are swept under the rug (a notable exception is \cite{Bladel}, which mentions the example of Lebesgue spine on page 136).

\smallskip

Solution to our Dirichlet problem exists if and only if each point of the boundary is \emph{regular} \cite{GT,Salsa}. If a component $\Omega_i$ is bounded we have at our disposal well-known existence results (e.g.~section 3.5.3 in \cite{Salsa}, sections 6.3 and 8.10 in \cite{GT}, section 6.3 in \cite{Evans} and broad discussion of problems on nonsmooth domains in \cite{Grisvard}). For the exterior region $\Omega_{\mathrm{e}}$ we have at least two strategies to prove the existence of solution. One is to combine geometric inversion in a sphere, which transforms unbounded to bounded domain, with the Kelvin transform \cite{AG,SW66,MM03}. The other strategy is to solve a sequence of problems on bounded, expanding domains \cite{MS60} and prove that sequence of this solutions converge to the solution on external domain via Harnack's inequality \cite{Moser61,GT,Evans}.

\smallskip

\begin{figure}[!ht]
\centering
\begin{tikzpicture}
\draw[lightgray,fill=lightgray] (0,0) [out=0,in=120] to (2,-0.5) [out=60,in=180] to (3,0.2) [out=0,in=180] to (4,0) -- (4,-1) -- (0,-1) -- (0,0);
\draw[very thick] (0,0) [out=0,in=120] to (2,-0.5) [out=60,in=180] to (3,0.2) [out=0,in=180] to (4,0); 
\draw[] (2,0.1) ellipse (0.2 and 0.1);
\draw[] (1.8,0.1) -- (2,-0.46) -- (2.2,0.1);
\draw[] (3,0.7) ellipse (0.2 and 0.1);
\draw[] (2.8,0.7) -- (3,0.2) -- (3.2,0.7);
\node at (3,-0.5) {\small $\Omega$};
\end{tikzpicture}
\caption{Illustration of the exterior cone condition.}
\label{fig:extcone}
\end{figure}
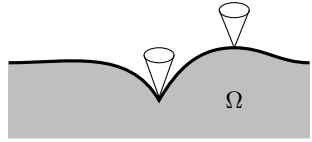

Now, regularity of boundary is rather impractical to check in practice, but there are various \emph{sufficient} conditions which imply the regularity of a boundary point. One of those is so-called \emph{exterior cone condition}, satisfied at the point $x \in \dd\Omega$ if there exists a closed circular cone $\Gamma_x$ with the vertex at $x$ and an open ball $B = B(x,r)$ with radius $r > 0$, such that $\cl{B} \cap \Gamma_x \subseteq (\rr^3 - \Omega)$ (see \cite{Salsa}, p.~145; see also remarks in \cite{Simon}, p.~231), as illustrated in Figure \ref{fig:extcone}. For example, exterior of a cube satisfies the exterior cone condition at each point of its boundary. Non-examples, worth having in mind, are sets with piecewise smooth boundaries which do not satisfy the exterior cone condition, such as 

\bc
$\big\{ (x,y) \in \rr^2 \mid y < \sqrt[4]{x^2} \, \big\}$ \ and \ $\big\{ (x,y,z) \in \rr^3 \mid z < \sqrt[4]{x^2 + y^2} \, \big\}$.
\ec

\medskip

\noindent
\emph{Uniqueness}. Again, as with the question of existence, we have two approaches to the proof. One is to apply maximum principles \cite{HL} on difference of solutions \cite{GT}. Note that on the unbounded, exterior region $\Omega_\mathrm{e}$ we rely on the additional assumption that solution $u$ asymptotically vanishes, $u(x) \to 0$ as $\norm{x} \to \infty$, which implies that $\max_{\cl{\Omega}_\mathrm{e}} |u(x)| = \max_{\dd\Omega_\mathrm{e}} |u(x)|$. For bounded $\Omega$ we have other, more elementary approach, based on the identity
\be\label{eq:idGreen}
\int_\Omega \hat{\epsilon}^{ab} (\nab{a}w)(\nab{b}w) \, \vol = \oint_{\dd\Omega} \hat{\epsilon}^{ab} w (\nab{a} w) n_b \, \df a - \int_\Omega w \nab{a} (\hat{\epsilon}^{ab} w \nab{b} w) \, \vol \ .
\ee
Assuming that the problem has two solutions, $u_1$ and $u_2$, we may look at $w = u_1 - u_2$ which again solves the differential equation (\ref{eq:ellip}) and vanishes on $\dd\Omega$, so that that both right hand terms in (\ref{eq:idGreen}) vanish. As by assumptions $\hat{\epsilon}^{ab}$ is positive definite, it follows that $w$ is constant on $\Omega$ and, due to boundary condition, $w = 0$ on $\cl{\Omega}$, implying the uniqueness.

\smallskip

\emph{Triviality}. A well known result implies that classical solutions of the Laplace equation are necessarily smooth. Not only that, but harmonic functions on nonempty open connected set $\Omega \subseteq \rr^3$ are even real analytic on that set (see e.g.~chapter 2.4 in \cite{GT} and chapter 3.3.7 in \cite{Salsa}). More general, for the linear elliptic partial differential equation with \emph{real analytic} coefficients, the solutions are necessarily real analytic (see \cite{MN57,Morrey58} and section 6.6 in \cite{Morrey}). We shall make use of this property in order to prove one important auxiliary result. Let us first recall an elementary lemma.

\smallskip

%%% https://mathoverflow.net/q/33614/
%%% https://math.stackexchange.com/q/2412111/
%%% Hopf maximum principle
%%% https://en.wikipedia.org/wiki/Hopf_maximum_principle
%%% Theorem 2.30 in Sørensen: Advanced Partial Differential Equations (notes by Marcel Schaub)

\blm
Let $\Omega \subseteq \rr^3$ be a nonempty open set, $f \in C^0(\cl{\Omega})$ and
\be
\int_B f \, \vol = 0
\ee
for any ball $B \subseteq \Omega$. Then $f = 0$ on $\Omega$.
\elm

\smallskip

\noindent
This lemma provides us with a simple test: if $u \in C^2(\Omega)$ satisfies
\be
\int_{\dd B} n_a \hat{\epsilon}^{ab} \nab{b} u \, \df a = 0
\ee
for any ball $B \subseteq \Omega$, then by the Stokes' theorem
\be
\int_B \nab{a} (\hat{\epsilon}^{ab} \nab{b} u) \, \vol = 0 \ ,
\ee
implying $\nab{a} (\hat{\epsilon}^{ab} \nab{b} u) = 0$ on $\Omega$.

\smallskip

\btm[generalization of Problem 2.2 in \cite{GT}] \label{tm:CauchyEllipt}

Let $u \in C^1(\cl{\Omega}) \cap C^2(\Omega)$ be solution of (\ref{eq:hardie}) with real analytic tensor $\hat{\epsilon}_{ab}$ on a basic connected open set $\Omega \subseteq \rr^3$ and $u = 0 = n_a \hat{\epsilon}^{ab} \nab{b} u$ on some open, smooth subset $S$ of $\dd\Omega$. Then $u = 0$ on whole $\Omega$.
\etm

\noindent
\emph{Proof}\footnote{The main idea for this particular version of proof in the special case of the Laplace equation originates from the online discussion at the StackExchange forum, \href{http://math.stackexchange.com/q/482576}{\texttt{http://math.stackexchange.com/q/482576}}}. Let $x \in S$ and $O_x$ an open connected neighbourhood of the point $x$, divided by $S$ in two connected components. We extend the function $u$ from $O_x \cap \cl{\Omega}$ to $O_x$, so that $\tilde{u}$ is zero on $O_x - \cl{\Omega}$. The extension $\tilde{u}$ is immediately of class $C^1$ on $O_x$. Now we claim that $\tilde{u}$ satisfies $\nab{a}(\hat{\epsilon}^{ab}\nab{b}\tilde{u}) = 0$ on $O_x$. We look at the integral of normal derivative $n_a \hat{\epsilon}^{ab} \nab{b} \tilde{u}$ over a boundary of balls in $O_x$. If such a ball is completely within $O_x \cap \Omega$ or $O_x - \cl{\Omega}$ then this integral is immediately zero. Otherwise, we must look more carefully at the integral,
\bal
\oint_{\dd B} \!\! n_a \hat{\epsilon}^{ab}\nab{b}\tilde{u} \, \df a = \int_{\dd B \cap \Omega} \!\! n_a \hat{\epsilon}^{ab}\nab{b} u \, \df a = \nonumber\\
= \int_{B \cap \Omega} \!\!\! \nab{a}(\hat{\epsilon}^{ab} \nab{b}u) \, \vol - \oint_{B \cap \dd\Omega} \!\!\! n_a \hat{\epsilon}^{ab}\nab{b} u \, \df a = 0 \ .
\eal
Whence, by the previous lemma, $\nab{a}(\hat{\epsilon}^{ab}\nab{b}\tilde{u}) = 0$ on $O_x$. Also, by classic results mentioned above, $\tilde{u}$ is also real analytic on $O_x$. Since $\tilde{u}$ is zero on an open subset of $O_x$ (and $O_x$ is by assumption connected), then $\tilde{u}$ is zero on whole $O_x$. Finally, we can extend this argument: Since $u = 0$ on $O_x \cap \Omega$, and $u$ is real analytic on $\Omega$, it follows that $u$ is zero on whole $\Omega$. \qed

\end{document}